\documentstyle[preprint,aps]{revtex}
\begin{document}
\draft
\title{\bf{\LARGE{Stationary 
Localized States Due to a
Nonlinear Dimeric Impurity Embedded in a Perfect 1-D Chain}}}
\author{B. C. Gupta and K. Kundu}
\address{Institute of physics, Bhubaneswar - 751 005, India}
\maketitle
\begin{abstract}
The formation of Stationary Localized states due to a nonlinear
dimeric impurity embedded in a perfect 1-d chain is studied here
using the appropriate Discrete Nonlinear Schr$\ddot{o}$dinger
Equation. Furthermore, the nonlinearity has the form, $\chi
|C|^\sigma$ where $C$ is the complex amplitude. A proper ansatz for
the Localized state is introduced in the appropriate Hamiltonian of
the system to obtain the reduced effective Hamiltonian. The
Hamiltonian contains a parameter, $\beta = \phi_1/\phi_0$ which 
is the ratio of stationary amplitudes at impurity sites. Relevant 
equations for Localized states are obtained from the fixed point 
of the reduced dynamical system. $|\beta|$ = 1 is always a 
permissible solution. We also find solutions for which $|\beta| 
\ne 1$. Complete phase diagram in the $(\chi, \sigma)$ plane 
comprising of both cases is discussed. Several critical lines 
separating various regions are found. Maximum number of Localized 
states is found to be six. Furthermore, the phase diagram 
continuously extrapolates from one region to the other.
The importance of our results in relation to solitonic solutions 
in a fully nonlinear system is discussed.
\end{abstract}
\narrowtext
\newpage
\section{Introduction}

Discrete Nonlinear Schr$\ddot{o}$dinger equation (DNLSE) is a set of 
$n$ coupled differential equations.
\begin{eqnarray}
i \frac{dC_{m}}{dt} &=& - \chi_{m} f_{m}(\mid C_{m} \mid) C_{m} 
+ V_{m,m+1} C_{m+1} + V_{m,m-1} C_{m-1} \nonumber \\
{\rm where}~~~~~~ V_{m,m+1} &=& V_{m+1,m}^{\star};~~ {\rm and}
~~m=1,2,3,.........n.
\end{eqnarray}
In eqn.(1) the nonlinearity appears through the functions 
$f_{m}(\mid C_m \mid)$ and $\chi_{m}$ is the nonlinearity parameter 
associated with the $m$-th grid point. Since, $\sum_{m} \mid C_m 
\mid^2$ is made unity by choosing appropriate initial conditions, 
$\mid C_m \mid^2$ can be interpreted as the probability of finding 
a particle at the $m$-th grid point. One way to derive this set of 
equations is to couple the vibration of masses at the lattice 
points of a lattice of $n$ sites to the motion of a quasi particle 
in the same lattice in the adiabatic approximation. 
The motion of the quasi particle is described,
however, in the framework of a tight binding Hamiltonian (TBH).
Same type of equation can also be obtained by nonlinear coupling of
anharmonic oscillators through both positions and momenta of the 
oscillators. The set of equations, thus derived, is called the 
discrete self-trapping equation (DST). These equations also posses 
a constant of motion analogous to $\sum_{m} \mid C_m \mid^2$ in the 
DNLSE. In fact, both the DST and the DNLSE contain the same number 
of constants of motion. However, in general the analytical solutions 
of eqn.(1) are not known. Numerous works, both analytical as well as 
numerical, on the DNLSE and the DST have been reported
\cite{a,b,c,d,e,f,g,h,i,j,k,l,m,n,o,p}.

One important feature of this type of nonlinear equations is that
these can yield stationary localized ( SL ) states and soliton--like
solutions. It has been shown that the presence of a nonlinear
impurity can produce SL states in one dimension. The first study 
was made using Green's function approach and authors considered 
$f(\mid C \mid)$ = $\mid C \mid^2$ \cite{15}. Later, one nonlinear 
impurity case has been generalized by taking $f(\mid C \mid) = \mid
C \mid^{\sigma}$ and formation of SL states studied in one, two 
and three dimensions \cite{16,17,18}. Furthermore, considering 
$f(|C|) = |C|^{\sigma}$, some discussions have been made about the
formation of SL states in the presence of two impurities in a 1-d
chain \cite{16}. In another development the Green function approach 
and the ansatz approach were synthesized to find SL states self 
consistently. Two types of nonlinear impurities, namely, 
$f(|C|) = |C|^{\sigma}$ where $\sigma$ is arbitrary and the 
rotational nonlinear impurity, embedded in linear hosts like 
1-d perfect chain and Caley tree have been considered \cite{bik}. 
The purpose of considering Caley
tree is to study the effect of connectivity on the formation of SL 
states. In case of nonlinear dimer, $|C_0|^2 = |C_1|^2$ ( probability
of the particle to stay at the nonlinear sites are same ) is assumed.
However, for $\sigma$ = 2, it was found that a $\chi_{cr}$ = 8/3
exists above which an extra SL state appears. In this SL state
$|C_0|^2 \ne |C_1|^2$. It is, therefore, imperative to study if 
such a solution occurs for all $\sigma$. This is the motivation 
behind this paper. The formation of SL states is studied here 
starting from a
Hamiltonian. The fixed point of the Hamiltonian \cite{mic} which 
generates the appropriate DNLSE can also produce the correct
equations governing the formation of SL states. Although this
approach is simpler, it however needs the proper ansatz. We further
note that the appropriate ansatz has been obtained in our earlier 
analysis \cite{bik}.

The organization of the paper is as follows. In sec.II we describe
the formalism part. In section III(A) the phase diagram and energy
diagram for the SL states satisfying $|\beta| \ne 1$ is presented.
In sec.III(B) we discuss the full phase diagram of SL states 
considering both the cases, namely, $|\beta| = 1$ and $|\beta| \ne
1$. In the last section we give a summary of our investigation.

\section {Formalism}

The eqn.(1) describing the system of 1-d chain consisting of a 
nonlinear dimer impurity of the kind $\chi |C|^{\sigma}$ can 
be derived from the Hamiltonian given by
\begin{equation}
H = \frac{1}{2} \sum_m (C_m^{\star} C_{m+1} + C_m C_{m+1}^{\star}) +
\frac{\chi}{\sigma +2} (|C_0|^{\sigma+2} + |C_1|^{\sigma+2}).
\end{equation}
A model derivation of the Hamiltonian is given in ref. \cite{bik}.
Since we are interested in the possible solutions for SL states,
we assume
\begin{eqnarray}
C_m &=& \phi_m exp(-iEt) \nonumber \\
{\rm where}~~~~~~~\phi_m &=& [sgn(E) \eta]^{m-1} \phi_1,~~~~~~~m \ge 1
\nonumber \\ 
{\rm and} ~~~~~~~~~\phi_{-|m|} &=& [sgn(E) \eta]^{|m|} \phi_0, ~~~~~~m
\le 0
\end{eqnarray}
Eqn.(3) is the exact form of $\phi_m$ in the presence of a dimeric 
nonlinear impurity and can be derived from Greens function analysis 
\cite{bik}. Here $0 < \eta < 1$ and is given by
\begin{equation}
\eta = \frac{\mid E \mid - \sqrt{E^2 - 4}}{2}.
\end{equation}
It is also to be noted that for $\chi$ = 0 all possible states of the
resulting linear system lies in the band defined by $\mid E \mid ~~
\le$ 2. Since eqn.(3) defines a localized state it will appear either
above the upper band edge or below the lower band edge of the linear
system depending on the sign of $\chi$. Introducing the $sgn(E)$ or
the signature of $E$ in eqn.(3) we take care of that possibility. We
further define $\beta = \phi_1 / \phi_0$ if $|\phi_1| \le |\phi_0|$.
Otherwise, we invert the definition of $\beta$. Because of the
symmetry in the system we shall get the same result. So, we
restrict $\beta$ in [-1,1]. Now from the normalization condition,
$\sum_m |C_m|^2$ = 1, we get
\begin{equation}
|\phi_0|^2 = \frac{1 - \eta^2}{1 + \beta^2}.
\end{equation}
Using eqn.(3), and eqn.(5) and the definition of $\beta$ we get an
effective Hamiltonian, $H_{eff}$ where
\begin{equation}
H_{eff} = \frac{\beta (1 - \eta^2)}{1 + \beta^2} + sgn(E) \eta +
\frac{\chi}{\sigma + 2} (1 - \eta^2)^{\sigma/2 + 1} (1 +
\beta^2)^{-(\sigma /2 + 1)} (1 + |\beta|^{\sigma + 2}).
\end{equation}
The Hamiltonian consists of two variables namely $\beta$ and $\eta$ 
because $\chi$ and $\sigma$ are constants. The stationary localized 
states correspond to fixed points of the reduced dynamical system 
described by $H_{eff}$. So to obtain the possible stationary localized
states we need solving two coupled algebraic equations in $\eta$ and
$\beta$ arising from setting \cite{k,mic}
\begin{equation}
\frac{\partial{H_{eff}}}{\partial{\beta}} = 0 ~~~~~~~{\rm and}~~~~~~~~ 
\frac{\partial{H_{eff}}}{\partial{\eta}} = 0.
\end{equation}
From the first relation of (7) we obtain 
\begin{equation}
(\beta^2 - 1)[\frac{(1 + \beta^2)^{\sigma/2}}{\beta \chi} - \frac{(1
- \eta^2)^{\sigma/2} (|\beta|^{\sigma} - 1)}{\beta^2 - 1}] = 0.
\end{equation}
We note that the term in the parentheses is finite for $|\beta| = 1$.
So, $|\beta|$ = 1 is always a solution of eqn.(8). In the other 
equation, namely, $\partial{H_{eff}}/\partial{\eta}$ = 0, if we set 
$|\beta|$ = 1, we obtain 
\begin{equation}
\frac{2^{\sigma /2}}{|\chi|} = \eta_{\pm} (1 \mp \eta_{\pm})^{-1}
(1 - \eta_{\pm}^2)^{\sigma /2}. 
\end{equation}
Here $\eta_{\pm}$ refers to the symmetric and antisymmetric cases
respectively. The phase diagram of SL states arising from eqn.(9)
has been discussed in detail in ref. \cite{bik}. We further note 
that in our previous analysis for $\sigma$ = 2, we obtained another 
$\chi_{cr} = \frac{8}{3}$ above which $|\beta| \ne$ 1. So, to explore 
the possibility of SL states in which $|\beta| \ne$ 1 for $\sigma$ 
other than 2, we set the term in the parentheses in eqn.(8) to zero. 
This then defines a relation between $\eta$ and $\beta$. Furthermore,
if we introduce the expression for $(1 - \eta^2)$, thus derived in 
the other equation of eqn.(7) we obtain
\begin{equation}
sgn(E) \eta = sgn(\beta) \frac{|\beta|^{-\sigma/2} -
|\beta|^{\sigma/2}}{|\beta|^{-(\sigma/2 + 1)} - |\beta|^{(\sigma/2 +
1)}}.
\end{equation}
\noindent We note that $\eta$ is a symmetric function of $\beta$ and 
$\frac{1} {\beta}$ and is always less than unity in magnitude
irrespective of the magnitude of $\beta$. Furthermore, when $|\beta| 
\rightarrow$ 1, we obtain $\eta = \pm sgn(E) \frac{\sigma}{\sigma +
2}$. At $\sigma$ = 2, $\eta = \frac{1}{2}$ and this has been obtained 
in our earlier work \cite{bik}. 
Since $\eta$ is by definition positive,
if $\beta < $ 0, we must have $E <$ -2 and vice versa. Again from 
the expression of $(1 - \eta^2)$ which should always be positive, 
it can be shown that for $\beta < 0 $ will imply $\chi < 0$ and 
vice versa. Hence, for $\beta < 0$, we must have $\chi < 0, E < 
-2$ and on the other hand for $\beta > 0$, we need $\chi > 0$ and 
$E > 2$. Consequently, by choosing $\chi$ positive we can restrict 
$\beta$ in [0,1]. We also note that $\beta \ne 1$ is not possible 
for the antisymmetric set and the analytical argument for this has 
been presented in ref. \cite{bik}. Now introducing eqn.(10) in 
the equation of $(1 - \eta^2)$ we obtain  
\begin{equation}
\frac{1}{\chi^{2/\sigma}} = \frac{(\beta^{-1} -
\beta) (\beta^{-\sigma/2} - \beta^{\sigma/2})^{2/\sigma}
(\beta^{-(\sigma + 1)} - \beta^{(\sigma +
1)})}{(\beta^{-1} + \beta) (\beta^{-1} - \beta)^{2/\sigma}
(\beta^{-(\sigma/2 + 1)} - \beta^{(\sigma/2 + 1}))^2} = F(\beta, 
\sigma)
\end{equation}

We note that we have $\beta \ge 0$. Furthermore, the right hand side 
of eqn.(11) is also a symmetric function of $\beta$ and $\beta^{-1}$ 
as like $\eta$. For $\sigma$ = 2, we find, from eqn.(11), $\chi_{cr} 
= \frac{8}{3}$ and this is consistent with the result obtained from 
Green's function analysis. So, to obtain the phase diagram of SL 
states for $\beta \ne$ 1, we analyze eqn.(11) and the results are 
discussed in the next section. 

\section{Results and discussions}
\noindent {\bf (A) Phase diagram and energy diagram of SL states
with $|\beta| \ne 1$} \\
We first note that the number of possible SL states is the number of
permissible solutions of eqn.(11) for $\beta \in$ [0,1]. So, we
need to know the behavior of $F(\beta, \sigma)$ as a function of
$\beta$ and $\sigma$. We see that for $\sigma$ = 2, $F(\beta,
\sigma)$ is monotonically increasing function of $\beta$ and this is
shown in fig.1. So, for $\sigma$ = 2 we can have no solution (no SL
state) or one solution (one SL state) of eqn.(11) depending on the
value of $\chi$. On the other hand for $\sigma$ = 5, we find a local
maximum in $F(\beta, \sigma$) and this is shown in fig.2.
Therefore, for this value of $\sigma$ we can atmost have two
solutions (two SL states) of eqn.(11). This clearly shows that there
is a critical value of $\sigma$ $(\sigma_{cr})$ above which $F(\beta,
\sigma)$ develops a local maximum. This $\sigma_{cr}$ can be
obtained from the minimum value of $\sigma$ satisfying the relation, 
$\partial{F(\beta, \sigma)}/\partial{\beta}$ = 0 for $\beta \in$
[0,1]. This is found graphically and  $\sigma_{cr}$ comes out to 
be $\approx$ 2.33. Therefore, for $\sigma < \sigma_{cr}$, since 
$F(\beta, \sigma)$ is monotonically increasing function of $\beta$, 
maximum value it takes for $\beta \in [0,1]$ is $F(1, \sigma)$.
Inasmuch as, the critical value of $\chi$ required to get a SL state
for this case is 
\begin{equation}
\chi_{cr} = \frac{2}{\sigma} [\frac{(\sigma + 2)^2}{2 (\sigma + 
1)}]^{\sigma/2}, 
\end{equation}
\noindent it implies that for $\sigma \le \sigma_{cr}$ we will have 
no SL states if $\chi < \chi_{cr}$ and only one SL state if $\chi 
\ge \chi_{cr}$.
On the other hand for $\sigma > \sigma_{cr}$ there will be two 
critical values of $\chi$ because $F(0, \sigma)$ = 0, $F(1, \sigma) 
\ne$ 0 and there is a local maximum of $F(\beta, \sigma)$ at some 
$\beta_0 \in [0, 1]$. Therefore, for $\sigma > \sigma_{cr}$, lower 
critical value of $\chi$ $(\chi_{lcr})$ to get a SL state is given 
by $\chi_{lcr} = [1/F(\beta_0, \sigma)]^{\sigma/2}$. The upper 
critical value of $\chi$ $(\chi_{ucr})$ separating the one and two 
SL states regions is same as that given in eqn.(12). Therefore, for 
$\sigma > \sigma_{cr}$, we will have no SL states if $\chi < 
\chi_{lcr}$, one SL state at $\chi = \chi_{lcr}$, two SL states 
for $\chi_{lcr} < \chi \ge \chi_{ucr}$ and again one SL
state above $\chi_{ucr}$. We further note that the upper critical
line for $\sigma > \sigma_{cr}$ joins smoothly with critical line for
$\sigma \le \sigma_{cr}$. These lines are shown by solid and dotted
lines respectively in fig.3. Salient features of the phase diagram
of SL states in the $(\chi, \sigma)$ as shown in fig.3 are discussed 
below.

(1) In fig.3 the region below the solid curve has 
no SL states. Every point of the blank region bound by the solid 
and the dotted curves represents only one SL state. The shaded region
bound by the solid and the dotted curves has two possible SL states.
(2) There are threshold values of $\chi$ and $\sigma$ below which no 
SL states appear. These values can be obtained from the relation 
$d\chi_{cr}/d\sigma$ = 0 where $\chi_{cr}$ is given in eqn.(12). 
Thus we obtain $\chi_{th1} \approx$ 2.593 and $\sigma_{th1} \approx$ 
1.645. This point is shown being surrounded by a small box 
in fig.3. There are also threshold values for both $\chi$ and
$\sigma$ below which there is no possibility of getting two SL
states. These threshold values are given by $\chi_{th2} = \chi_{lcr}
\approx 2.866$ and $\sigma_{th2} = \sigma_{cr}$ = 2.33 as discussed 
in earlier paragraph. This point is shown by a star in the figure. 

Therefore, in the region bounded by $\chi \in$ $[\chi_{th1}$, 
$\chi_{th2}$] and $\sigma \in$ [$\sigma_{th1}$, $\sigma_{th2}$] we 
will have only one SL state. Consequently, for any $\chi$ between 
$\chi_{th1}$ and $\chi_{th2}$ there will be two critical values of 
$\sigma$, namely, $\sigma_1$ and $\sigma_2$. These points are shown 
in the figure for $\chi = 2.7 \in [\chi_{th1}, \chi_{th2}]$. For this
value of $\chi \in$ [$\chi_{th1}$, $\chi_{th2}$] there will be no 
SL state below $\sigma_1$, one SL state between $\sigma_1$ and 
$\sigma_2$ and again there will be no SL states above $\sigma_2$. 
For a fixed value of $\chi > \chi_{th2}$ there will be three 
critical values of $\sigma$. For example, for $\chi$ = 4.25 $> 
\chi_{th2}$, these three values of $\sigma$ are 0.585, 3.29 and 4.28, 
respectively. These are denoted by $\sigma_3$, $\sigma_4$ and 
$\sigma_5$, respectively in the figure. Now for the fixed $\chi$ = 
4.25, there will be no SL states below $\sigma = 0.585$, one SL state 
for $0.585 < \sigma < 3.29$, two SL states for $3.29 < \sigma < 
4.28$, one SL state at $\sigma$ = 4.28 and again no SL state above 
$\sigma = 4.28$. We also notice that one SL state appears on the 
dotted line as well as on the solid line for $\sigma \le \sigma_{cr}$ 
and these correspond to the case where $\beta$ = 1. 
We further note that at $\sigma = 0$., the system contains a linear 
dimeric impurity with each site having the site-energy, $\chi$. So, 
to obtain a state fully localized on the dimer, 
but symmetric ( $\beta = 1$ ) we need infinite
magnitude for $\chi$. Precisely for this reason along the solid 
line for $\sigma \le \sigma_{cr}$ in fig.3 $\chi \rightarrow 
\infty$ as $\sigma \rightarrow 0$. On the other hand, as $\sigma$ 
goes to infinity, the dimer site-energies 
will go to zero  and a perfect system will be obtained. So, this 
cluster localized state will vanish. This in turn implies that       
$\chi \rightarrow \infty$ as $\sigma \rightarrow 0$.  So, along the 
critical line where $\beta = 1$, there will be another $\sigma_{cr}$,
{\em i. e.}, $\sigma_{th1}$ in 
our earlier discussion, for which $\chi_{cr}$ in eqn.(12) will assume
the minimum value, {\em i. e.}, $\chi_{th1}$. Similarly, the solid 
line for $\sigma > \sigma_{cr}$ represents one cluster localized SL 
state in which two impurity sites have different amplitude.

Here we consider variation of the energy of SL states with $\chi$ for
$\sigma <$ 2.33 as well as $\sigma >$ 2.33. The energy of the SL
states can be calculated using eqn.(4). Fig.4 shows the energy 
of the SL states as a function of $\chi$ for $\sigma$ = 2. It is
clear from the figure that the SL state starts appearing at $\chi$ 
= 8/3 and lies above the upper band edge of the perfect system. We 
see that the energy of the SL state increases almost linearly with
$\chi$. That means as $\chi$ increases, the localization length of
the state decreases and hence localization becomes stronger with
the increase of $\chi$. In fig.5 we have plotted the energy of
the SL states as a function of $\chi$ for a fixed value of $\sigma 
= 4 > 2.33$. Here also no state, two states and one state regions get
reflected as in the fig.2. It is clear that for $\chi <$ 4.025
there is no energy of SL state, at $\chi = 4.025$ there is one energy
of SL state, for $4.025 < \chi \le 6.489$ there are two energies of SL
states for each $\chi$ and above $\chi = 6.489$ there is again one 
energy of SL state for each $\chi$. All possible states appear above 
the upper band edge. As $\chi$ increases from 4.025
to 6.489, energy of one of the states increases with $\chi$ and that 
of the other state decreases towards the band edge state. So, as 
$\chi$ increases from 4.025 to 6.489, one of the state localizes 
strongly while localization of the other state becomes weaker. 
After $\chi$ = 6.489, one state disappears and energy of the other 
state continues to increase with the increase of $\chi$. 

We now consider the variation of energy with $\sigma$ by first
considering  critical lines. Along the upper critical line 
(solid curve in fig.3), 
$\beta$ = 1 and $\eta = \sigma/(\sigma + 2)$. This is true for 
both $\sigma > \sigma_{cr}$, and $\sigma \le \sigma_{cr}$. 
So, as $\sigma \rightarrow \infty, \eta
\rightarrow 1$. Therefore, the energy of the SL states along this
line will go towards the upper band edge as $\sigma \rightarrow
\infty$. SL states that appear above this critical line also have
this property. On the other hand, along the lower critical line 
(dotted curve in fig.3)
$\beta_0 \rightarrow 0$ as $\sigma \rightarrow \infty$ which
can be shown from the movement of the maximum of $F(\beta, \sigma)$
as a function of $\sigma$. So, along this critical line, $\eta
\rightarrow 0$ as $\sigma \rightarrow \infty$. Consequently $E
\rightarrow \infty$. Therefore, the energy of one of the SL states in
the shaded region of phase diagram will increase as $\sigma$
increases. The energy of the other SL state will probably decrease.
We now consider a specific example.

Fig.6 shows how the energy of SL state or states vary with $\sigma$
for fixed value of $\chi$. The example that we choose is $\chi = 5 >
\chi_{th2}$. We have plotted for a range of $\sigma$ between 2 and 6.
Here $\sigma_{cr}$ = 3.58. We see that for $\sigma < 3.58$ only one 
state appear above the upper band edge and the energy
of the state decreases with increasing $\sigma$. At $\sigma = 3.58$
an another state ( shown by dotted line in the figure ) appears 
with lower energy above the upper band edge of the perfect 
system. The energy of
the new state increases with the increase of $\sigma$. The energy 
of the first state goes on decreasing as $\sigma$ increases and 
hence the localization of the state becomes weaker. But localization 
of the new state becomes stronger as $\sigma$ increases. Ultimately  
energies of both the states join at certain value of $\sigma$ (5.618)
and then they disappear as expected (see fig.3). The disappearance of
the SL states can be understood from the fact that as $\sigma$
increases for a fixed $\chi$, the effective nonlinearity at the
impurity sites decreases and hence the system approaches towards the
perfect system.

\noindent {\bf (B) The full phase diagram of SL states}\\
Taking into account all possible solutions for $\beta$ =1 as well as
$\beta \ne$ 1 we will have the complete phase diagram. This is shown
in fig.7. This figure shows several separated regions and the number
of SL states range from zero to six.
First we consider the region for $\sigma <$ 2.
There are four separated regions on the left side of $\sigma$ = 2
line. The region I contains only one SL state and this comes from  
the symmetric set corresponding to $\beta$ =1. In region II, there 
are three SL states. Two of them are from the antisymmetric set 
corresponding to $\beta$ = 1 and the other one is the contribution 
from  the symmetric set. Region III has two SL states, one from      
the symmetric set with $\beta$ = 1 and the other is for $\beta \ne$
1. The region marked by IV has four SL states. Two of them from
the antisymmetric set, one from the symmetric set for $\beta$ =1
and the last one is from the case where $\beta \ne$ 1. Along the
$\sigma$ =2 line there are three critical values of $\chi$ and they
are marked each by a star. These values are 1, 8/3 and 8 respectively
and have been discussed in our earlier work \cite{bik}. Now we look 
at the regions on the right side of $\sigma = 2$ line. The region V 
contains two SL states and both are contributed from the symmetric 
set. Region VI contains no SL state. The region VII contain          
two SL states and both of them are from the solutions when 
$\beta \ne$ 1. The region VIII has altogether three SL states, two of
them are contributed from the symmetric set for $\beta$ = 1 and the
other one comes from the solution for $\beta \ne 1$. Region IX
contains five SL states, two of them come from the symmetric set, 
two from the antisymmetric set and the last one
arises from the case where $\beta \ne 1$. Region X and the triangular
region bounded by regions V, VI, VII and VII contain four SL states.
Two of them arise from the symmetric set and the other two come
form the solution for $\beta \ne$ 1. There is a region containing the
maximum number of SL states and that region occurs for large value of
$\chi$. The line separating region VIII, IX and the line
separating the regions VII, X will cross each other at a larger value
of $\chi$ and will produce a region of six SL states. Each of
the symmetric, the antisymmetric set and the solutions for $\beta 
\ne$ 1 contributes two SL states in this region. It is to be noticed 
that there is continuity throughout the phase diagram as we go from 
one region to the other. We also find that if there are N (N is even)
SL states in a region, energies of N/2 states increase and that of 
N/2 states decrease as $\chi$ is increased.
On the other hand for odd N, energies of (N + 1)/2 states
increase and that of (N - 1)/2 states decrease with increasing
$\chi$.

\section{Conclusion:}

The formation of Stationary Localized states due to a dimeric
impurity in a perfect 1-d linear system has been studied using the
DNLSE. The nonlinearity considered has the form, $\chi |C|^{\sigma}$ 
where $\chi$ and $\sigma$ are arbitrary. In our previous work we used
Green's function approach and the phase diagram for two particular 
situations was obtained. However from the previous analysis we were 
able to find the exact analytical structure of SL states. This 
enabled us to find the effective Hamiltonian, $H_{eff}$ corresponding
to the DNLSE considered here. 
This Hamiltonian is a function of two variables
$\eta$ and $\beta$. The equations for SL states are obtained from the
fixed points of the dynamical system described by $H_{eff}$.

The total phase diagram of SL states then consists of two situations.
In one case $|\beta| = 1$ and in the other case $|\beta| \ne 1$. We
also show that the signs of $\beta$ as well as $\chi$ determine the
position of SL states. For example, for $\beta > 0$ and $\chi > 0$ 
which has been considered here, SL states appear above the upper band
edge of the linear system. It is important to note that the phase 
diagram is quite rich in structure. It contains several $\sigma_{cr}$
and $\chi_{cr}$ to get different number of SL states. The maximum 
number of SL states we obtain is six. Since the analysis is complete,
we conclude that the maximum number of 
possible SL states for the system
considered here is six. It is well established that for one nonlinear
impurity, we can get atmost two SL states. On the other hand, here we
have six. Therefore it is worthwhile to find out the relation between
the number of possible SL states and the number of impurities.
Furthermore, the study of SL states from nonlinear clusters of
similar type is important for understanding the formation of solitons
in totally nonlinear system. For example, the present dimer problem
is important for analyzing the formation of solitons peaking in
between lattice sites. This aspect will be discussed elaborately 
elsewhere.

\begin{figure}
\caption {$F(\beta, \sigma)$ and $1/\chi$ is plotted as a
function of $\beta$ in the range between 0 and 1 for
a fixed value of $\sigma$ = 2.} 
\end{figure}

\begin{figure}
\caption {$F(\beta, \sigma)$ is plotted
as a function of $\beta$ in the range between 0 and 1 for a fixed 
value of $\sigma$ = 5. $\beta_0$ is the position of maximum of 
$F(\beta, \sigma)$}
\end{figure}

\begin{figure}
\caption {The phase diagram of SL states for $\beta \ne 1$ is shown.
Here $\sigma_1$ = 1.645, $\sigma_2$ = 2.33, $\sigma_3$ = 0.585, 
$\sigma_4$ = 3.29 and $\sigma_5$ = 4.28. The point surrounded by 
small box and the point marked by star has 
coordinates (2.593, 1.645) and (2.866, 2.33) respectively.}
\end{figure}

\begin{figure}
\caption {Variation of energy of SL state as a function of $\chi$
for $\sigma = 2$ is shown. Vertical line touches the $\chi$ axis at
8/3 and is drawn to show the
critical value of $\chi$ to get SL state.}
\end{figure}

\begin{figure}
\caption {Variation of energy of SL states as a function of
$\chi$ for a fixed $\sigma = 4$ is shown. Vertical lines are drawn to
show the critical values of $\chi$ separating no SL state, two SL
states and one SL state regions. The dotted and the dashed vertical
lines touch the $\chi$ axis at 4.025 and 6.489 respectively.}
\end{figure}

\begin{figure}
\caption {This shows the variation of energy of SL states as a
function of $\sigma$ for a fixed value of $\chi = 5$. Vertical 
lines are drawn to show the transition points. The small and 
long vertical lines touch the $\sigma$ axis at 3.58 and  5.618 
respectively.}
\end{figure}

\begin{figure}
\caption {This shows the full ( both the case, namely, $\beta = 1$
and $\beta \ne 1$ is taken in to account ) phase diagram in the 
$(\chi, \sigma)$ plane. There are several marked regions containing
different number of SL states.}
\end{figure}
\end{document}